# Dial-a-ride problem with modular platooning and en-route transfers

**Zhexi Fu, Joseph Y. J. Chow***
C2SMART University Transportation Center, NYU Tandon School of Engineering
*Corresponding author: joseph.chow@nyu.edu

## Abstract

Modular vehicles (MV) possess the ability to physically connect/disconnect with each other and travel in platoon with less energy consumption. A fleet of demand-responsive transit vehicles with such technology can serve passengers door to door or have vehicles deviate to platoon with each other to travel at lower cost and allow for en-route passenger transfers before splitting. A mixed integer linear programming (MILP) model is formulated to solve this "modular dial-a-ride problem" (MDARP). A heuristic algorithm based on Steiner-tree-inspired large neighborhood search is developed to solve the MDARP for practical scenarios. A set of small-scale synthetic numerical experiments are tested to evaluate the optimality gap and computation time between exact solutions of the MDARP using commercial software and the proposed heuristic. Large-scale experiments are conducted on the Anaheim network with 378 candidate join/split nodes to further explore the potentials and identify the ideal operation scenarios of MVs. The results show that MV technology can save up to 52.0% in vehicle travel cost, 35.6% in passenger service time, and 29.4% in total cost against existing on-demand mobility services in the scenarios tested. Results suggest that MVs best benefit from platooning by serving "enclave pairs" as a hub-and-spoke service.

**Keywords:** Modular Vehicle, Dial-a-ride Problem, Vehicle Platooning Problem, Variable Capacity, Synchronized En-route Transfers, Pickup and Delivery Problem with Transfer

# 1. Introduction

Conventional mass transit and demand-responsive mobility systems use vehicles with fixed capacity and cannot adapt effectively to the temporal and spatial demand variations with a satisfactory level of service (Dakic et al., 2021). During peak hours and in areas with a high volume of demand requests, low-capacity vehicles may exacerbate passenger wait time and in-vehicle time due to detours. On the contrary, operating high-capacity vehicles may lead to low vehicle occupancy and unnecessary energy consumption during off-peak hours and in low demand density areas. As for demand-responsive transit (DRT) services like paratransit, their low efficiency in service throughput (e.g, each vehicle only serving several passengers at a time) may cause more traffic congestion problems in the city. There is no existing transportation modal technology that can adapt to both high and low spatial-temporal demand density scenarios.

The emerging modular vehicle (MV) technology, such as NEXT Future Transportation (2022), may be able to address the mismatch problems between heterogeneous demand and fixed vehicle capacity mentioned above (Guo et al., 2018; Caros and Chow, 2020; Chen et al., 2019; Chen et al., 2020; Dakic et al., 2021). The MV technology allows vehicles to connect and disconnect with each other in motion so that (1) they can join and travel in a platoon for less energy consumption, and (2) to reposition on-board passengers between vehicles (en-route transfer) at the same time. With these two distinct advantages over such existing microtransit systems as Via and MOIA, MVs can expand their vehicle capacity according to the demand during peak hours and also separate individually to provide on-demand services during off-peak hours or at lower density areas. Guo et al. (2018) showed the value of having such flexibility in adjusting vehicle size, while Caros and Chow (202) found that it can save on both operation cost and user disutility over current MOD systems in simulation tests using demand data from Dubai, depending on the operational structure. An illustrative diagram is shown in **Figure 1** to demonstrate the operation of MVs.

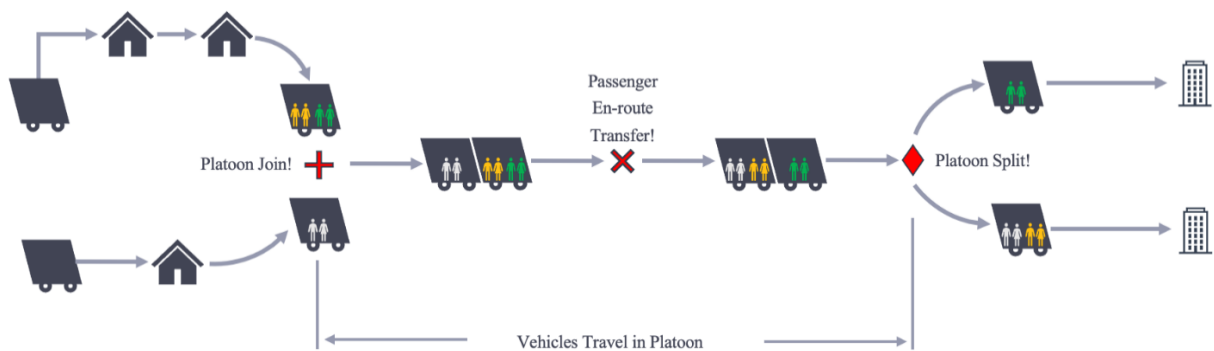

Figure 1. Modular vehicle diagram: platoon join/split, passenger en-route transfer.

There are three major challenges to overcome to optimize the demand-responsive routing of these vehicles. First, since MVs may join and leave a platoon at any location and time, tracking the status of each vehicle in both spatial and temporal dimensions is necessary to ensure the synchronization of the docking and undocking process of a platoon. For vehicles that travel in the same platoon, their departure time at the preceding location and arrival time at the following location should be identical. Next, passengers could be relocated and transferred between MVs while they travel in platoons. The second challenge is to search and identify passenger en-route transfers that further improve the total cost. The third challenge comes with the changes in platoon size. A longer platoon with more MVs can be regarded as a single new bus-like platform with expanded capacity. However, if any vehicle leaves the platoon, the on-board carrying capacity of this platoon varies accordingly. Thus, the variable capacity is an important feature and a challenge of MVs.

While Fu and Chow (2022) proposed a method to route vehicles considering synchronized spatial-temporal transfers, it does not handle the benefits and challenges from vehicle platooning. In this study, we propose a mixed integer linear programming (MILP) model for a dial-a-ride problem with modular platooning, or a "modular dial-a-ride problem" (MDARP). The MILP formulation tracks the vehicle platoon status, captures the passenger en-route transfer, and address the variable capacity feature of MV platoons. To solve large-scale MDARPs, we propose a heuristic based on Steiner-tree-inspired large neighborhood search to construct, search, and improve MV platoons from an existing non-platooning routing solution.

## 2. Literature review

An illustrative example is presented first to demonstrate the differences and benefits in the operation of modular vehicles against existing mobility-on-demand services. Then, we provide a comprehensive literature review on relevant studies in the past. We define the MDARP. As introduced in previous section, modular vehicles can be physically connected with each other to travel in platoon with zero following gaps, which leads to reduction in air resistance and energy consumption. Platoon savings can be further made under an automated vehicle fleet setting where drivers are not needed for each vehicle (Tirachini and Antoniou, 2020). Meanwhile, passengers with similar destinations can be relocated and transferred between the connected MVs to optimize their delivery paths. Thus, the MDARP is a complex combination and extension of multiple sub-problems, such as the dial-a-ride problem with transfers (DARPT) and the vehicle platooning problem (VPP). To the best of our knowledge, the MDARP is a new problem that has not been studied in the literature.

### *2.1 Problem illustration*

This section demonstrates the problem settings and potential benefits of using MVs through an illustrative example. Two operation policies are considered: 1) a solo (S) mode (the non-platooning operation) and 2) a modular (M) mode. The solo mode operates as a conventional DRT service where passengers board a vehicle at a pickup location and alight at a destination location without any transfers or platooning. In this case, each individual vehicle is regarded as an independent unit and the capacity of each vehicle cannot be combined. As for the modular mode, vehicles are allowed to join and connect as a platoon and thus move together to save vehicle travel cost, in terms of fuel savings from less air drag. Moreover, connected MVs allow passengers to transfer en-route, i.e. be relocated between platooned vehicles such that requests with similar destinations can be grouped together for delivery after splitting.

Consider 3 requests (each with a paired pickup and drop-off locations) and 2 vehicles on a 24-node undirected graph (each link can go on both directions) shown in **Figure 2**. This is a coarser representation of an infrastructure network, where each node is a pickup or drop-off location, or a candidate platoon join or split location, and links are shortest paths that do not include any of those nodes. The pick-up locations of 3 requests are at nodes {8, 4, 5}, and their corresponding drop-off locations are at nodes {20, 19, 24}. Vehicles are initially located at nodes {1, 6}, without any specified depot. For the simplicity of the problem illustration, each request consists of 2 passengers and each vehicle has a maximum capacity of 4 passengers on-board. We assume that all requests and vehicles enter the system at time $t = 0$ (in-system time) and the time used for the platoon join/split operation is ignored (assumed to be zero). Except links (9,15) and (10,16), the travel costs on each link are set to 1.

The objective is to minimize the weighted sum of vehicle travel cost and passenger service time (objective weights of vehicle travel cost and passenger service time are both set to 1 in this example without any loss of generality). The vehicle travel cost for each vehicle can be reduced by $\eta = 10\%$ for each additional platooned vehicle along the platooned paths. The passenger service time is measured by the time difference between the passenger drop-off time and in-system time, which includes the passenger wait time and in-vehicle travel time. In addition, since vehicles may arrive at the platoon join/split location at different times (spatial-temporal synchronization), the passenger service time also accounts for this possible platoon join/split delay time.

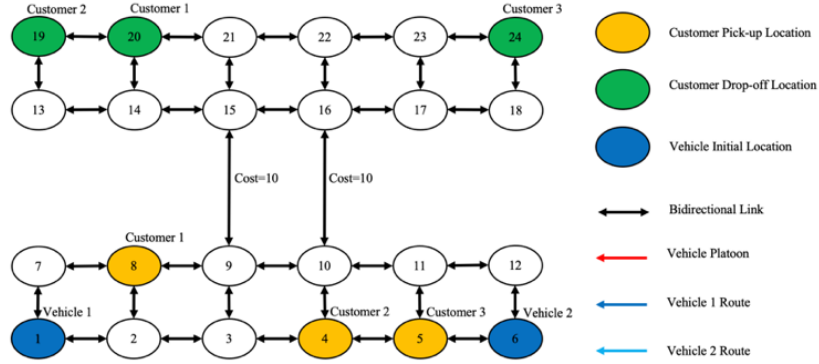
Figure 2. Illustrative example: initial locations.

For the solo mode, the optimal solution can be observed as shown in **Figure 3(a)**. Vehicle 1 is assigned to serve request 1 only, whereas vehicle 2 is assigned to serve requests 2 and 3. The solution has a total cost of 140.

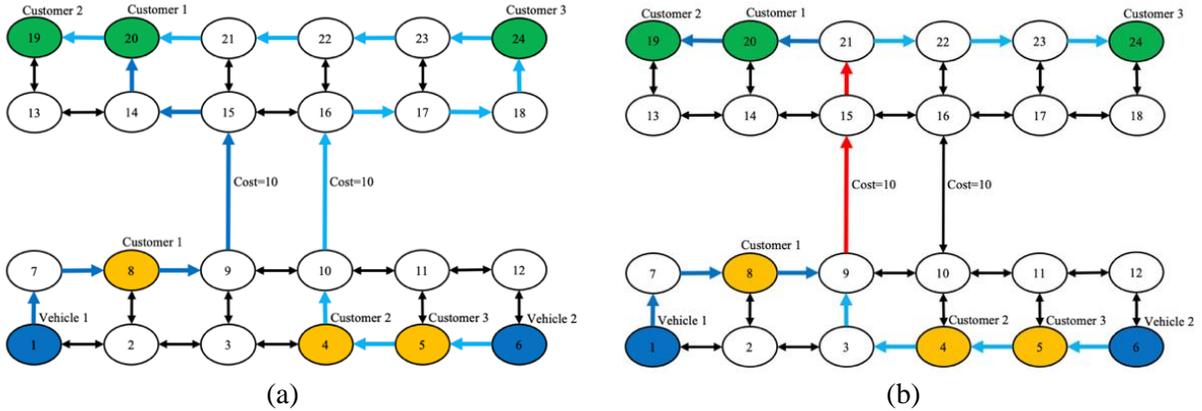
Figure 3. Routes for (a) S1, (b) S2.

In the modular mode, the solution is shown in **Figure 3(b)**. Vehicle 1 is assigned to pick up request 1, and vehicle 2 is assigned to pick up request 2 and 3, which is the optimal assignment for request pickups. The vehicles travel in platoon through nodes [9, 15, 21]. Vehicle 1 arrives at the platoon join node {9} earlier than vehicle 2, which causes a time delay to customer 1. The platoon join delay time impacts the passengers carried by the vehicle who arrives first. Along the path where vehicles travel in platoon, request 2 is transferred from vehicle 2 to vehicle 1. It is assumed that transfers can be made within the length of a link. After the platoon split, vehicle 1 completes the delivery of request 1 and 2, whereas vehicle 2 only finishes the delivery of request 3. The results are summarized in **Table 1**. The total cost equals to 133.8 in the modular operation, which is better than the optimal solution in solo mode. The relative cost difference between solo (S) and modular (M) mode is calculated resulting in total cost reductions of 4.43% against the solo mode.

Table 1. Summary of illustrative example results

| Operation Mode | Assignment | Vehicle Travel Cost | Passenger Service Time | Total Cost |
|---|---|---|---|---|
| Solo (S) | Veh 1 → Req 1<br>Veh 2 → Req 2, 3 | 36 | 104 | 140* |
| Modular (M) | Veh 1 → Req 1<br>Veh 2 → Req 2, 3 | 31.8<br>(-11.67%) | 102<br>(-1.92%) | 133.8*<br>(-4.43%) |

|  | *Pla* over {9, 15, 21} |  |  |  |

Notes: 1) *Veh* for vehicle, *Req* for request, *Pla* for platoon
     2) * for optimal solution in each operation mode
     3) percentages shown in () are cost differences against solo mode, calculated as $\frac{M-S}{S} \times 100\%$

## 2.2 Prior research

There are several studies on the impacts and efficiencies of vehicle platoons. Tsugawa et al. (2016) showed the effectiveness of platooning in energy saving and transportation capacity due to short gaps between vehicles through truck platooning experiments and projects. Nguyen et al. (2019) focused on the ride comfort and travel delay. Sethuraman et al. (2019) studied the bus platooning effects in an urban environment. Since the type of adjacent vehicle influences the energy saving of each vehicle, Lee et al. (2021) focused on the platoon formation strategy with heterogeneous vehicles and found that a bell-shaped platoon pattern is generally effective for energy saving.

The vehicle platooning problem (VPP) has been investigated as a network flow problem where flows may be assigned to seemingly longer paths between origin-destination (OD) pairs because they can platoon with other vehicles flows to increase cost savings. Larsson et al. (2015) developed a MILP formulation for the truck platooning problem and presented several heuristic algorithms for large-scale instances to minimize the total fuel consumption. With the same objective, Luo et al. (2018) proposed a coordinated platooning MILP model that integrates the speed selection and platoon formation/dissolution into the problem formulation. They also proposed a heuristic decomposed approach and tested on a grid network and the Chicago-area highway network. Boysen et al. (2018) investigated an identical-path truck platooning problem to explore various aspects that could impact the efficiency of platoons, such as the platoon formation process, inter-vehicle distance, maximum platoon length and tightness of time windows. Bhoopalam et al. (2018) presents a comprehensive overview on relevant truck platooning studies. However, all these previous literatures only focus on the minimization of vehicle energy consumption within the network flow optimization problem. There is no consideration of user costs involving the pickup and delivery of requests as a vehicle routing problem.

The combination of VPP with DARP requires searching for the join/split locations and maximizing the platoon paths and savings, which is a more complex extension of DARP. The presence of platooning means that a complete graph structure (i.e. conventional structure used in vehicle routing problems) would not work because platooning decisions depend on knowing path proximities to each other.

The second major challenge in the modular vehicle routing problem comes with the en-route transfer feature, where passengers could freely move and relocate between connected vehicles for more flexible and optimized routings. MDARP, which includes en-route transfers, can be seen as a variant of the DARPT or, more broadly, pickup and delivery problem with transfers (PDPT). A PDPT makes transfers at inactive stops or hubs, where passengers finish the entire transfer movement at a specific location. Cortés et al. (2010) considered transfers at a pre-specified static location, where each transfer node $r$ is split into two separate nodes, a start node $s(r)$ and finish node $f(r)$. They formulated a link-based MILP model and proposed a branch-and-cut solution method based on Benders Decomposition to solve the problem. Rais et al. (2014) proposed a MILP model for the PDPT problem at candidate transfer locations with and without time windows. However, their model does not explicitly track the vehicle time along its route and thus is not able to optimize the time used for transfer. Fu and Chow (2022) extended the model from Rais et al. (2014) and proposed a set of new constraints that can track the vehicle arrival time at every stop along its path. A vehicle is allowed to wait at transfer locations for passenger transfers within a maximum time limit. However, modeling of en-route transfers *within platoon vehicles in motion* has not been addressed in the literature in the context of DARP.

Third, since MVs can be assembled together in a platoon and treated as an entire new vehicle with expanded capacity, traditional capacity constraints might not be applicable for the operation of MVs. The on-board capacity of a platoon is the sum of all carrying limits of MVs in that platoon. Passengers served by a platoon of MVs belong to the entire platoon rather than individual vehicles. As a consequence, the platoon capacity changes dynamically with the join and split of vehicles, which requires constantly tracking the spatial and temporal status of vehicles. The literature on this variable capacity feature has been discussed in transit operations. For example, Chen et al. (2019, 2020) proposed both discrete and continuous methods that considers the variable capacity used in the operation of modular vehicles. However, their study only focuses on the joint design of dispatch headway and vehicle capacity on fixed route shuttle systems. Dakic et al. (2021) proposed an optimization model for the flexible bus dispatching system that jointly determines the optimal configuration of modular bus units and their dispatch frequency at each bus line. They use a three-dimensional macroscopic fundamental diagram (3D-MFD) to capture the dynamics of traffic congestion and complex interactions between the modes at the network level, assuming that the MVs impact the congestion level and the study area is homogenous. Tian et al. (2022) studied the optimal planning of public transit services with modular vehicles by determining the optimal location and capacity of stations where modular units can be assembled and dissembled. They formulate the problem as a mixed-integer nonlinear program (MINLP) and apply surrogate model-based optimization approaches for large-size problems. Similar studies (Zhang et al., 2020; Liu et al., 2021; Shi and Li, 2021; Wu et al., 2021; Li and Li, 2022) also focus on the operation of modular vehicles for public transit services. Additionally, Lin et al. (2022) present a paradigm for the bi-modality feature of modular vehicles, which integrates transit services with last-mile logistics to serve both passenger and freight with the same fleet.

So far, all relevant studies on the innovative MV technology focus on either vehicle platooning problem, pickup and delivery problem with transfers, or MV with variable capacity on public transit services, separately. This study integrates the vehicle platooning problem with request pickup and delivery, considers passenger en-route transfers during vehicle platooning, and addresses the variable capacity feature of platoons all together. The key literature of MDARP is summarized in **Table 2**.

Table 2. Summary of key literature

| **Vehicle platooning problem** | | |
|---|---|---|
| *Study* | *Methodology* | *Major Contributions* |
| Larsson et al. (2015) | MILP model and two-phase heuristic algorithm | Considered the truck platooning problem with routing, speed-dependent fuel consumption, and platooning formation/split decisions. |
| Boysen et al. (2018) | MILP model and efficient algorithms | Presented a basic scheduling problem for the platoon formation process along a single path. |
| Luo et al. (2018) | MILP model and decomposed heuristic approach | Integrated and solved multiple speed selections, scheduling, routing, and platoon formation/dissolution into coordinated platooning problems. |
| **Pickup and delivery problem with transfers** | | |
| *Study* | *Methodology* | *Major Contributions* |
| Cortés et al. (2010) | MILP model and branch-and-cut solution method based on Benders Decomposition | Formulated the PDPT problem with a static transshipment facility and splitable transfers. |
| Rais et al. (2014) | MILP model | Formulated the PDPT problem with a number of candidate transfer locations and time window constraints on directed networks. |
| Fu and Chow (2022) | MILP model and two-phase heuristic algorithm | Proposed a new formulation for the PDPT problem to optimize the temporal and spatial synchronization of transfers. |
| **Modular vehicles** | | |
| *Study* | *Methodology* | *Major Contributions* |

| Chen et al. (2019) | MILP model and dynamic programming algorithm | Used discrete modeling method to jointly design the dispatch headways and vehicle capacities of MAVs. |
| Dakic et al. (2021) | Methodological framework | Jointly optimize the configuration of modular bus units and their dispatch frequency at each bus line. |
| Tian et al. (2022) | MINLP with linearization and surrogate model based optimization approaches | Location and capacity optimization of the docking/undocking stations for new modular-vehicle transit services. |

In summary, to fill the research gaps in the concept of MV technology, we propose a MILP model for the MDARP and a heuristic algorithm to solve it. The major contributions of our paper are summarized as follows:

1) We formulate a mathematical model for the dial-a-ride problem that integrates vehicle platooning with request pickup and delivery, considers passenger en-route transfers during vehicle platooning, and addresses the variable capacity feature of platoons at the same time.
2) A heuristic based on Steiner tree-inspired local neighborhood search for join/split locations is proposed to solve the MDARP for large-scale problems.
3) We conduct a set of small- and large-scale numerical experiments to validate the feasibility of MVs. Results show that using MVs can save up to 52% in vehicle travel cost, 36% in passenger service time, and 29% in total cost against existing MOD services, depending on the operational setting.

## 3. Mathematical model

We now present our proposed MILP model for MDARP. Given a fleet of vehicles with their initial locations and a set of requests with their pick-up and drop-off locations, the objective is to find the optimal dispatch assignments of vehicles to requests and corresponding routes at minimum total cost. The total cost is measured by the weighted sum of vehicle travel cost and passenger service time. Vehicles are allowed to operate in platoon and thus save the vehicle travel cost from lower air resistance. The passenger service time is calculated by the difference between the request drop-off time and in-system time, which includes passenger wait time, in-vehicle travel time and possible platoon delay time.

The model requires the use of an undirected graph instead of a complete graph structure because of the need to quantify proximities of vehicle paths for platooning. There are different options to pursue in that direction. There are multicommodity flow formulations of vehicle routing problems on a directed graph (e.g. Garvin et al., 1957; Letchford and Salazar-González, 2015) including those on a time-expanded network (e.g. Mahmoudi and Zhou, 2016). The latter discretizes time into intervals. Two previous studies (Rais et al., 2014; Fu and Chow, 2022) instead have continuous time arrival but the design on an undirected graph prevents cyclic vehicle routes.

For example, suppose there are two customers with pickup and drop-off at {2,4} and {3,5}, with an initial vehicle location at node 1, on a graph shown in **Figure 4(a)**. A vehicle would not be able to pass twice at node 3. In order to let the vehicle travel the same location repetitively, we employ a multi-layer network structure by adding duplicate layers of network, as shown in **Figure 4(b)**. It has the advantage of the time-expanded network without the restriction of discretized time and only requires layers as needed instead of duplicating one layer every time interval. Layer 1 includes nodes {1,2,3,4,5} whereas layer 2 includes nodes {1',2',3',4',5'}. Each set of duplicate nodes (e.g., node 1 and 1') are connected from the upper layer to the lower layer with zero cost. Each duplicate node is equivalent to the original node on the network. For example, the customer request (2,4) can be picked up at either node 2 or node 2' and dropped off at either node 4 or node 4', but not at both duplicate nodes at the same time. Thus, with the same example on our modified multi-layer network structure, the optimal vehicle route in terms of node number becomes 1→3→2→2'→3'→4'→5', which achieves the multi-visit feature with the same optimal cost.

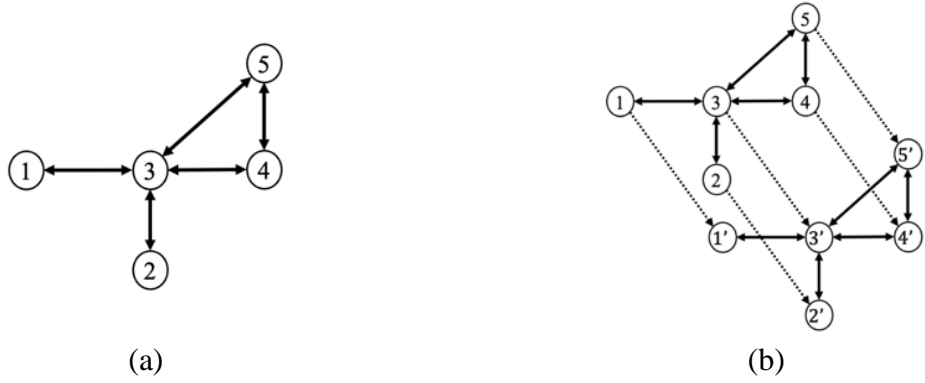
Figure 4. (a) Original undirected network (b) modified multi-layer network

## 3.1 Basic MILP formulation

The MDARP is defined on an undirected graph $G(N, A)$ over an operational time horizon $[0, T]$. $N$ is a set of nodes, consisting of vehicle initial and destination locations, request pickup and drop-off locations, and platoon join/split locations. $A$ is the set of undirected arcs, where $A_i^+$ and $A_i^-$ represent the set of inbound and outbound arcs at node $i$, respectively. The notation is shown in **Table 3**.

Table 3. Model notations

| Notations | Definitions |
| --- | --- |
| *Parameters* | |
| $d_{ij}$ | travel distance for arc $ij \in A$ |
| $\tau_{ij}$ | travel time for arc $ij \in A$ |
| $L$ | number of network layers |
| $K$ | set of vehicles ready for service |
| $s_k$ | starting location of vehicle $k \in K$ |
| $e_k$ | ending location of vehicle $k \in K$ |
| $c_k$ | capacity of vehicle $k \in K$ |
| $t_k^V$ | ready time for service of vehicle $k \in K$ |
| $R$ | set of customer requests |
| $o_{rl}$ | pick-up location of request $r \in R$ on layer $l \in L$ |
| $d_{rl}$ | drop-off location of request $r \in R$ on layer $l \in L$ |
| $q_r$ | number of passengers in request $r \in R$ |
| $t_r^Q$ | in-system time of request $r \in R$ |
| $u$ | maximum platoon length |
| $\eta$ | platoon cost saving rate |
| $M$ | large positive constant |
| | |
| *Decision variables* | |
| $X_{ijk}$ | 1 if vehicle $k$ traverses arc $ij$, and 0 otherwise |
| $Y_{ijkr}$ | 1 if vehicle $k$ carries request $r$ onboard and traverses arc $ij$, and 0 otherwise |
| $S_{kri}$ | 1 if vehicle $k$ is assigned to pick up and/or drop off request $r$ at node $i$, and 0 otherwise |
| $F_{rikm}$ | 1 if request $r$ transfers from vehicle $k$ to vehicle $m$ at node $i$, and 0 otherwise |
| $T_{ik}^V$ | time at which vehicle $k$ arrives at node $i$ |
| $T_{ir}^Q$ | time at which request $r$ arrives at node $i$ |
| $T_{ik}^U$ | dwell time of vehicle $k$ at node $i$ |
| $P_{ijkm}$ | 1 if vehicle $k$ travels in platoon with vehicle $m$ on arc $ij$, and 0 otherwise |
| $P_{ijk}^N$ | number of vehicles that travel in platoon with vehicle $k$ on arc $ij$ |

*Dummy variables*

| | |
|---|---|
| $Z_{ijk}$ | non-negative dummy variable for ensuring the continuity of $T_{ik}^V$ |
| $U_{ijk}$ | non-negative dummy variable for ensuring the continuity of $T_{ik}^U$ |
| $W_{kri}$ | non-negative dummy variable for measuring the pick-up and/or drop-off time of $S_{kri}$ |
| $C_{ijkm}$ | non-negative dummy variable that captures the total number of passengers carried by vehicle $m$ when vehicle $k$ and $m$ travel in platoon on arc $ij$ |

Let $K$ be the set of vehicles available to serve customers at time $t_k^V \in [0,T]$. For a vehicle $k \in K$, we use $s_k$ and $e_k$ to denote its starting and ending locations. The vehicle ending location is assigned to a dummy depot, which is set the same as the initial location/depot for a static problem, or assigned to a designated zone when idle under a dynamic reoptimization setting. We assume by default that the travel cost from any node to $e_k$ is zero without loss of generality to leave out $e_k$ from the graph. This way, implementing this model as a reoptimization as part of an online algorithm is possible with some modifications to assign the idle vehicle to a new designated location. Let $R$ be the set of customer pick-up and drop-off requests. For a customer request $r \in R$, $q_r$ denotes the number of passengers for the request, $o_{rl}$ denotes the pick-up location on network layer $l \in L$, and $d_{rl}$ denotes the corresponding drop-off location on network layer $l \in L$. The basic MILP model is shown in Eq. (1) – (9).

$$\text{Min: } \alpha \sum_{k \in K} \sum_{(i,j) \in A} d_{ij}(X_{ijk} - \eta P_{ijk}^N) + \beta \sum_{r \in R} q_r \left( \sum_{l \in L} T_{d_{rl}r}^Q - t_r^Q \right) \tag{1}$$

Subject to

$$\sum_{(i,j) \in A_i^-} X_{ijk} = 1, \quad \forall k \in K, i = s_k \tag{2}$$

$$\sum_{(j,i) \in A_i^+} X_{jik} = 1, \quad \forall k \in K, i = e_k \tag{3}$$

$$\sum_{(i,j) \in A_i^-} X_{ijk} - \sum_{(j,i) \in A_i^+} X_{jik} = 0, \quad \forall k \in K, \forall i \in N \setminus \{s_k, e_k\} \tag{4}$$

$$\sum_{l \in L} \sum_{k \in K} \sum_{(i,j) \in A_i^-} Y_{ijkr} - \sum_{l \in L} \sum_{k \in K} \sum_{(j,i) \in A_i^+} Y_{jikr} = 1, \quad \forall r \in R, i = o_{rl} \tag{5}$$

$$\sum_{l \in L} \sum_{k \in K} \sum_{(j,i) \in A_i^+} Y_{jikr} - \sum_{l \in L} \sum_{k \in K} \sum_{(i,j) \in A_i^-} Y_{ijkr} = 1, \quad \forall r \in R, i = d_{rl} \tag{6}$$

$$\sum_{k \in K} \sum_{(i,j) \in A_i^-} Y_{ijkr} - \sum_{k \in K} \sum_{(j,i) \in A_i^+} Y_{jikr} = 0, \quad \forall r \in R, \forall i \in N \setminus \{o_{rl}, d_{rl}\} \tag{7}$$

$$\sum_{k \in K} Y_{ijkr} \leq 1, \quad \forall r \in R, \forall (i,j) \in A \tag{8}$$

$$X_{ijk} \in \{0,1\}, \quad \forall k \in K, \forall (i,j) \in A \tag{9}$$

$$Y_{ijkr} \in \{0,1\}, \quad \forall k \in K, \forall r \in R, \forall (i,j) \in A \tag{10}$$

$$S_{kri} \in \{0,1\}, \quad \forall k \in K, \forall r \in R, \forall i \in \{o_{rl}, d_{rl}\} \tag{11}$$

$$F_{rikm} \in \{0,1\}, \quad \forall r \in R, \forall i \in N, \forall k, m \in K, k \neq m \tag{12}$$

$$T_{ik}^V \geq 0, \quad \forall k \in K, \forall i \in N \tag{13}$$

$$T_{ir}^Q \geq 0, \quad \forall r \in R, \forall i \in N \tag{14}$$

$$T_{ik}^U \geq 0, \quad \forall k \in K, \forall i \in N \tag{15}$$

$$P_{ijkm} \in \{0,1\}, \quad \forall k, m \in K, k \neq m, \forall (i,j) \in A \tag{16}$$

$$P_{ijk}^N \in [0, u-1], \quad \forall k \in K, \forall (i,j) \in A \tag{17}$$

$$Z_{ijk} \geq 0, \quad \forall k \in K, \forall (i,j) \in A \tag{18}$$

$$U_{ijk} \geq 0, \quad \forall k \in K, \forall (i,j) \in A \tag{19}$$

$$W_{kri} \geq 0, \quad \forall k \in K, \forall r \in R, \forall i \in \{o_{rl}, d_{rl}\} \tag{20}$$

$$C_{ijkm} \geq 0, \quad \forall k, m \in K, k \neq m, \forall (i,j) \in A \tag{21}$$

Objective function (1) minimizes the total cost of vehicle travel cost and passenger service time, weighted by constant values $\alpha$ and $\beta$ in a general form. The exact amounts of weights $\alpha$ and $\beta$ need to be calibrated. The cost savings from vehicle platooning are deducted by a rate of $\eta$ from the vehicle travel cost, as shown in the first term of objective function. We assume that as the platoon length increases, the average travel cost of each vehicle decreases. The second term of the objective function measures the passenger service time, which is calculated by the difference between the passenger drop-off time and in-system time. Thus, the passenger service time includes the wait time before pickup, in-vehicle travel time and possible platoon delay time. Constraints (2) and (3) ensure that each vehicle leaves its initial location and ends its trip at the ending location. If a vehicle is not used for service, it will be assigned to the ending location without any cost. Constraints (4) maintain the vehicle flow conservation at any node except its initial and ending locations. Constraints (5) and (6) ensure that each request is picked up and dropped off by exactly one vehicle on the multi-layer network. Constraints (7) maintain the passenger flow conservation at any node except its pick-up and drop-off locations. Note that we assume all vehicle initial locations are placed on the top network layer and then vehicles traverse the network to serve requests. However, requests might be picked up and dropped off by different vehicles on different network layers, which is the reason that constraints (5) and (6) consider over all vehicles and network layers. Constraints (8) enforce each request to be served by only one vehicle at a time. Integral and non-negativity constraints are defined in Eqs. (9) to (21).

### 3.2 *Continuity of vehicle arrival time constraints*

MVs need to physically connect and disconnect with each other to form a platoon, which requires spatial and temporal synchronization between vehicles. Moreover, vehicles might arrive at the platoon join location at different times and wait for other vehicles, causing extra delay to passengers on-board. To ensure the spatial-temporal synchronization between vehicles and support the optimization of passenger service time, it is necessary to track the exact travel time of each vehicle along the path.

Linearized constraints (22) – (29) were proposed and proven to be equivalent in Fu and Chow (2022) to ensure the continuity of vehicle arrival time along the path. For each vehicle $k \in K$, it can start the service at its earliest available time $t_k^V \in [0, T]$ from the initial location $s_k$. The arrival time at each following location along the vehicle path equals to the sum of arrival time and dwell time at previous location and the travel time between them.

$$T_{ik}^V \geq t_k^V, \quad \forall k \in K, i = s_k \tag{22}$$

$$Z_{ijk} - Z_{jik} = 0, \quad \forall k \in K, \forall (i,j) \in A \tag{23}$$

$$Z_{ijk} \leq (X_{ijk} + X_{jik})M, \quad \forall k \in K, \forall (i,j) \in A \tag{24}$$

$$Z_{ijk} \leq T_{ik}^V + \tau_{ij} X_{ijk} + U_{ijk}, \quad \forall k \in K, \forall (i,j) \in A \tag{25}$$

$$Z_{ijk} \geq T_{ik}^V + \tau_{ij}X_{ijk} + U_{ijk} - [1 - (X_{ijk} + X_{jik})]M, \quad \forall k \in K, \forall (i,j) \in A \tag{26}$$

$$U_{ijk} \leq X_{ijk}M, \quad \forall k \in K, \forall (i,j) \in A \tag{27}$$

$$U_{ijk} \leq T_{ik}^U, \quad \forall k \in K, \forall (i,j) \in A \tag{28}$$

$$U_{ijk} \geq T_{ik}^U - (1 - X_{ijk})M, \quad \forall k \in K, \forall (i,j) \in A \tag{29}$$

Constraints (22) ensure that vehicles can only start the operation after their earliest available time $t_k^V \in [0, T]$. Constraints (23) – (29) are a linearization of constraints to ensure that the vehicle arrival time is consistent along its path.

### 3.3  Proposed passenger pickup and drop-off time constraints

The passenger pickup and drop-off time is measured by the arrival time of the assigned pickup and delivery vehicle. Under certain circumstances, other vehicles may also traverse the same pickup and delivery location of a specific request, so it is necessary to identify the designated pickup and delivery vehicle for each request. Therefore, constraints (30) and (31) are proposed to capture the customer-vehicle pickup and delivery assignments, respectively. The decision variable $S_{kri} = 1$ if vehicle $k$ is assigned to pick up or drop off request $r$ at node $i$, and 0 otherwise.

$$S_{kri} = \sum_{(i,j) \in A_i^-} Y_{ijkr} - \sum_{(j,i) \in A_i^+} Y_{jikr}, \quad \forall k \in K, \forall r \in R, \forall i \in \{o_{rl}\} \tag{30}$$

$$S_{kri} = \sum_{(j,i) \in A_i^+} Y_{jikr} - \sum_{(i,j) \in A_i^-} Y_{ijkr}, \quad \forall k \in K, \forall r \in R, \forall i \in \{d_{rl}\} \tag{31}$$

Constraints (32) – (35) are proposed as linearized constraints to measure the corresponding pickup and delivery time for each customer request. $W_{kri}$ is the dummy variable for the product of $S_{kri}T_{ik}^V$, the customer-vehicle assignment multiplied by the vehicle arrival time.

$$T_{ir}^Q = \sum_{k \in K} W_{kri}, \quad \forall r \in R, \forall i \in \{o_{rl}, d_{rl}\} \tag{32}$$

$$W_{kri} \leq S_{kri}M, \quad \forall k \in K, \forall r \in R, \forall i \in \{o_{rl}, d_{rl}\} \tag{33}$$

$$W_{kri} \leq T_{ik}^V, \quad \forall k \in K, \forall r \in R, \forall i \in \{o_{rl}, d_{rl}\} \tag{34}$$

$$W_{kri} \geq T_{ik}^V - (1 - S_{kri})M, \quad \forall k \in K, \forall r \in R, \forall i \in \{o_{rl}, d_{rl}\} \tag{35}$$

### 3.4  Proposed vehicle platoon constraints

For any two vehicles $k$ and $m$ ($k \neq m$) traveling in a platoon over arc $ij$, their departure time at node $i$ and arrival time at node $j$ are guaranteed to be the same by constraints (36) – (39). The decision variable $P_{ijkm} = 1$ if vehicle $k$ and $m$ travel in platoon over the arc $ij$, and 0 otherwise. Constraints (40) ensure that $P_{ijkm} = 1$ if and only if both vehicles travel on the same arc $ij$.

$$(T_{ik}^V + T_{ik}^U) - (T_{im}^V + T_{im}^U) \leq M(1 - P_{ijkm}), \quad \forall k, m \in K, k \neq m, \forall (i,j) \in A \tag{36}$$

$$(T_{im}^V + T_{im}^U) - (T_{ik}^V + T_{ik}^U) \leq M(1 - P_{ijkm}), \quad \forall k, m \in K, k \neq m, \forall (i,j) \in A \tag{37}$$

$$T_{jk}^V - T_{jm}^V \leq M(1 - P_{ijkm}), \quad \forall k, m \in K, k \neq m, \forall (i,j) \in A \tag{38}$$

$$T_{jm}^V - T_{jk}^V \leq M(1 - P_{ijkm}), \quad \forall k, m \in K, k \neq m, \forall (i,j) \in A \tag{39}$$

$$2P_{ijkm} \leq X_{ijk} + X_{ijm}, \qquad \forall k, m \in K, k \neq m, \forall (i,j) \in A \qquad (40)$$

Constraints (41) are used to capture the number of vehicles traveling with vehicle $k$ on the arc $ij$, while constraints (42) limit the maximum allowed number of vehicles in a platoon.

$$P_{ijk}^N \leq \sum_{m \in K, m \neq k} P_{ijkm}, \qquad \forall k \in K, \forall (i,j) \in A \qquad (41)$$

$$P_{ijk}^N \leq u - 1, \qquad \forall k \in K, \forall (i,j) \in A \qquad (42)$$

## 3.5 Proposed passenger en-route transfers constraints

As one of the key features of MV, passengers can be relocated between vehicles when they are traveling in the same platoon. For any platoon of MVs traveling on the arc $ij$, we assume that passengers can be transferred at any time when they travel from node $i$ to node $j$. This differs from Fu and Chow (2022), where passengers can only transfer at an inactive node. Since passenger flows are captured at each node in our proposed MILP model, here we set the decision variable $F_{rikm} = 1$ if request $r$ is transferred from vehicle $k$ to vehicle $m$ at node $i$, and 0 otherwise. Constraints (43) are used to identify the passenger en-route transfers. Thus, for any request $r$ transferred from vehicle $k$ to $m$ over the arc $ij$, we would have either $F_{rikm} = 1$ or $F_{rjkm} = 1$. Furthermore, constraints (44) ensure that the passenger en-route transfer between vehicles $k$ and $m$ over the arc $ij$ can only happen when they travel in platoon through node $i$ or $j$.

$$\sum_{(j,i) \in A_i^+} Y_{jikr} + \sum_{(i,j) \in A_i^-} Y_{ijmr} \leq F_{rikm} + 1, \qquad \forall r \in R, \forall i \in N, \forall k, m \in K, k \neq m \qquad (43)$$

$$F_{rikm} \leq \sum_{(i,j) \in A_i^-} P_{ijkm} + \sum_{(j,i) \in A_i^+} P_{jikm}, \qquad \forall r \in R, \forall i \in N, \forall k, m \in K, k \neq m \qquad (44)$$

## 3.6 Proposed variable capacity constraints

Two challenges exist in formulating the capacity constraints of MVs. First, during the operation, each individual MV may join and leave the platoon at any time and location, which leads to frequent changes of the on-board carrying capacity. This requires the model to track the status of each vehicle, whether it is traveling in platoon or by itself. Second, since MVs can physically form a longer bus-like platform, a platoon consisting of multiple vehicles should be regarded as an entire unit. Under certain circumstances, a platoon of MVs can serve more requests than the same number of individual vehicles (higher service throughput), considering requests with more than one individual.

Traditional vehicle capacity constraints in the literature only limit the number of carried passengers on each individual vehicle, which is not suitable to address the above-mentioned challenges. In our model, constraints (45) are proposed as linearized constraints to accommodate the variable changes in platoon length and capacity.

$C_{ijkm}$ is equivalent to $P_{ijkm}(\sum_{r \in R} q_r Y_{ijmr})$ and shown in constraints (46), which is the number of passengers carried by vehicle $m$ if vehicle $k$ and vehicle $m$ travel in platoon on the same arc $ij$. The left-hand side of constraints (45) indicates the total number of passengers in the platoon that vehicle $k$ involves, while the right-hand side of constraints (45) is the total on-board capacity of the platoon. If vehicle $k$ travels by itself on arc $ij$, constraints (45) becomes the traditional vehicle capacity constraints as $\sum_{r \in R} q_r Y_{ijkr} \leq c_k X_{ijk}$. In addition, constraints (45) also ensure that a request served by any vehicle can only happen when that vehicle also travels on the same arc. Constraints (47) – (49) are the linearization constraints for the dummy variable $C_{ijkm}$.

$$\sum_{r \in R} q_r Y_{ijkr} + \sum_{m \in K, m \neq k} C_{ijkm} \leq c_k X_{ijk} + \sum_{m \in K, m \neq k} c_m P_{ijkm}, \qquad \forall k \in K, \forall (i,j) \in A \qquad (45)$$

$$C_{ijkm} = P_{ijkm}\left(\sum_{r\in R} q_r Y_{ijmr}\right), \qquad \forall k,m \in K, k \neq m, \forall (i,j) \in A \tag{46}$$

$$C_{ijkm} \leq P_{ijkm}M, \qquad \forall k,m \in K, k \neq m, \forall (i,j) \in A \tag{47}$$

$$C_{ijkm} \leq \sum_{r\in R} q_r Y_{ijmr}, \qquad \forall k,m \in K, k \neq m, \forall (i,j) \in A \tag{48}$$

$$C_{ijkm} \geq \sum_{r\in R} q_r Y_{ijmr} - (1 - P_{ijkm})M, \qquad \forall k,m \in K, k \neq m, \forall (i,j) \in A \tag{49}$$

### *3.7 Proposed problem variants and extensions*

**Hard time windows.** Throughout this study, we try to solve the optimal routes for the MDARP and explore the potentials of MVs by penalizing the request service time in the objective function without applying hard time windows for pickups and deliveries. However, we do have constraints to ensure that a request can only be served after their in-system time. Since we have provided constraints to measure the passenger service times, constraints (50) and (51) can be applied if hard time windows are required for pickup and delivery.

$$a_r^o \leq T_{ir}^Q \leq b_r^o, \qquad \forall r \in R, \forall i \in \{o_{rl}\} \tag{50}$$

$$a_r^d \leq T_{ir}^Q \leq b_r^d, \qquad \forall r \in R, \forall i \in \{d_{rl}\} \tag{51}$$

**Platoon saving rate.** In contrast to traditional truck platooning studies in literature, MVs possess lighter mass and are physically connected with zero inter-vehicle gap between each other. In a platoon with three trucks (Tsugawa et al., 2016), the leading vehicle saves the least fuel from platooning (up to about 9% with 5m gap), while the following vehicles in the middle of platoon benefit the most from energy savings (up to about 23% with 5m gap), while the fuel saving of the tail vehicle does not change much regardless of the inter-vehicle gap (about 15%). Since the design of MVs expects 0m gaps in the platoons, we assume that the average energy consumption among all MVs decreases at similar or better rates as the platoon length increases because of economies of scale. However, in this study, we do not distinguish the vehicle position in platoon, such as the leading vehicle, following vehicle, and tail vehicle. Instead, we average the cost savings among all vehicles for the operator cost.

The vehicle travel cost term of objective function shown in Section 3.1 can be modified and substituted by equations (52), where $\eta_1$ represents an initial platoon saving rate if platooning occurs and $\eta_2$ represents additional saving bonus with more vehicles in platoon. A new decision variable, $P_{ijk}^V$, is defined in Eq. (53) to indicate whether vehicle $k$ travels in platoon on arc $ij$ or not.

$$\alpha \sum_{k\in K}\sum_{(i,j)\in A} d_{ij}(X_{ijk} - \eta_1 P_{ijk}^V - \eta_2(P_{ijk}^N - 1)) \tag{52}$$

$$\sum_{m\in K, m\neq k} P_{ijkm} \leq P_{ijk}^V M, \qquad \forall k \in K, \forall (i,j) \in A \tag{53}$$

In summary, the full MILP model for MDARP consists of Eqs. (1) – (45) and (47) – (49). As this simplifies to a DARP when the set of vehicle platoon, passenger en-route transfers and variable capacity constraints are relaxed from the model, the problem is already NP-hard and thus requires efficient heuristics to solve problems of practical size.

## 4. Proposed heuristic algorithm

Since the MDARP can be simplified to a DARP without vehicle platooning and passenger transfers, which is known to be NP-hard, we propose a route improvement heuristic based on modifying feasible solo (S) mode (non-transfer non-platoon) solution (which can be obtained using any existing DARP algorithm).

Based on the routes from solo mode, we iteratively seek to improve them with modular (M) mode solutions. For convenience, we use the word "platoon" to describe feasible MV routes, where a platoon might consist of multiple MVs with different origins and destinations but at least share one common path. The main idea is to partition the general problem into multiple subproblems. In Section 4.1, the solo mode routes are deconstructed and then reconstructed between pairwise individual vehicles to find feasible two-vehicle platoons. In Section 4.2, we first iteratively merge between feasible MV platoons. If there is any new platoon created, we continue to explore merging until no platoons can be joined with each other. Then, remaining individual vehicles are iteratively inserted into platoons found previously to extend the common platoon paths and maximize the cost savings. Thus, the heuristic consists of two major parts: 1) a Steiner-tree-inspired neighborhood search algorithm to modify the solo mode routes and find two-vehicle MV platoons, and 2) an improvement heuristic to iteratively merge between feasible MV platoons and then insert individual vehicles to platoons to maximize the common platoon path and save more cost.

Unlike the MILP model using link costs on an undirected graph to find the optimal routes, we use pre-processed shortest-path travel distance (SPD) and time (SPT) matrices between pairwise nodes *as a complete graph* in the heuristic. As a result, the multi-layer structure from the MILP model is no longer needed in our proposed heuristic algorithm. In addition, the output vehicle routes only need to be expressed in terms of starting and ending vehicle locations, request pickup/drop-off locations, and platoon join/split locations, rather than all explicit nodes along the vehicle route. Among the two variants in Section 3.7, only the linear platoon rate is considered (Eqs. (52) – (53)).

### 4.1 Two-vehicle platoon

**Algorithm 1** is used to search for two-vehicle platoons and passenger en-route transfers between individual vehicles. The main idea of searching for platoon join/split location was inspired from the Steiner Tree problem, especially the case with four terminals and two Steiner points (Beasley, J., 1992; Zachariasen, M., 1999; Bhoopalam et al., 2018). Based on the large neighborhood search (LNS) algorithm, **Algorithm 1** iteratively destroys and reconstructs the solo mode vehicle routes to search for platoon possibilities. Passengers can be re-assigned to a different vehicle and their pickup and drop-off sequences can be changed as well. However, since MVs traveling in platoon can enlarge the on-board carrying limit, vehicle capacity constraint is temporarily ignored when reconstructing the routes.

Next, for any pair of reconstructed routes, we check every segment along the new vehicle routes, search for potential platoon join and split locations, force them to divert and travel in platoon between the join and split nodes, and then calculate the corresponding cost. If the platoon deviation satisfies capacity constraints and time window constraints, we continue to iteratively extend the shared common platoon path in the routes to maximize the platoon length and cost savings.

Last, we destroy and rebuild the delivery paths for passengers considering en-route transfers within those platoons.

**Algorithm 1.** Two-vehicle platoon formation

| | |
|---|---|
| Input: | Graph $G(N, A)$, set of $R$ with $o_r, d_r, q_r$ and $t_r^Q = a_r^o$, set of $K$ with $s_k$ and $c_k$. Platoon saving rate $\eta$. $\varsigma = 1$. |
| Initialization: | SPD matrix $D_{ij}$ and SPT matrix $T_{ij}$ for any nodes $i, j \in N$. Solo mode routes $P^S$ and costs $C^S$. |
| 1. | **For** any $k, m \in K, k \neq m$ **do** |
| 2. |     Deconstruct solo mode paths $p_k^S, p_m^S \in P^S$ into sets of stops. |
| 3. |     Randomly reconstruct new sets of routes as $P_{km}^M$, ignoring $c_k, c_m$ and $t_r^Q$. |
| 4. |     **For** each segment in routes $p_k^M, p_m^M \in P_{km}^M$ **do** |
| 5. |         Search for $J$ and $S$ from $N$ and to generate candidate platoons $P_{km}^C$ (*Algo 2*). |
| 6. | Calculate new costs of $P_{km}^C$ with $\eta$, while satisfying $c_k$, $c_m$ and $t_r^Q$. |
| 7. | **For** each $p_{km}^C \in P_{km}^C$ **do** |
| 8. | **While** $\varsigma = 1$ **do** |
| 9. | Search for additional $J$ prior to start and $S$ after end of existing platoon (*Algo 2*). |
| 10. | Calculate new costs with modified routes. |
| 11. | Record the extension if improvement found, otherwise $\varsigma = 0$. |
| 12. | Deconstruct extended $p_{km}^C$ after end of platoon into sets of stops. |
| 13. | Randomly reconstruct new sets of passenger delivery routes between $k, m$ as $\psi$. |
| 14. | Calculate new costs of $\psi$ with passenger en-route transfers. Set $\varsigma = 1$. |
| 15. | Rank and select platoons by descending order of savings against $C^S$ as $p_k^M, p_m^M \in P^M$ with associated $c_k, c_m \in C^M$. Remove selected $k, m \in K$ from $P^S$ and $C^S$. |
| Output: | Two-vehicle modular mode routes $P^M$ and costs $C^M$, updated $P^S$ and $C^S$. |

Note: *Algo* for Algorithm

The ideal platoon join and split locations may already exist in current vehicle routes or in the neighborhood along the path, which requires a neighborhood search algorithm in the latter case. As shown in **Figure 5(a)**, given two route segments $j_1 \to s_1$ and $j_2 \to s_2$, both $j_1$ and $j_2$ are considered as candidate platoon join locations, while both $s_1$ and $s_2$ are considered as candidate split locations.

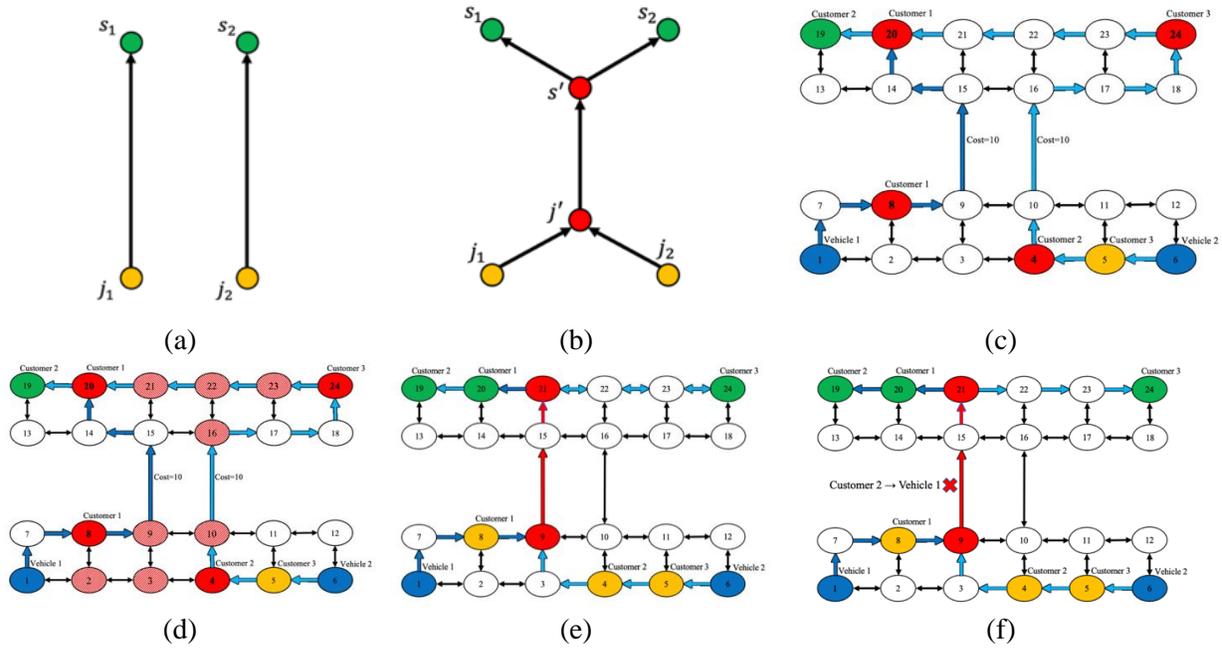

Figure 5. Platoon join/split location search illustration.

We then use a neighborhood search method, as presented in **Algorithm 2**, to find and return a number of potential join ($J$) and split ($S$) nodes in the neighborhood (**Figure 5(b)**). The ideal platoon join/split nodes should be located in-between the candidate join/split nodes with minimum connecting distance and corresponding difference to reduce the vehicle detour cost as much as possible. The cost coefficient $\varphi$ is used to make trade-offs between these two measurements. For convenience, we set the value of $\varphi = 1$ throughout this study. Next, we iterate over each combination of candidate platoon join and split nodes $j' \in J, s' \in S$, to divert the vehicle routes and force them to travel in platoon along the segment $j' \to s'$. We calculate the new cost of modified vehicle routes to identify the maximum feasible platoon length between $j'$ and $s'$.

**Algorithm 2.** Platoon join/split location search

| | |
|---|---|
| Input: | Nodes $n_1, n_2 \in N$, SPD matrix $D_{ij}$. |
| | Maximum number of returned nodes $N_{max}$, and cost coefficient $\varphi$. |
| Initialization: | Cost $\delta_i = M$ for $i \in N$. |
| 1. | **For** each $i \in N\backslash\{n_1, n_2\}$ **do** |
| 2. | Record and update the cost $\delta_i = (D_{n_1 i} + D_{n_2 i}) + \varphi|D_{n_1 i} - D_{n_2 i}|$. |
| 3. | Rank and select $N_{max}$ nodes by descending orders of $\delta_i$ as $J$ or $S$. |
| Output: | Candidate platoon join locations $J$ or split locations $S$. |

We use the same illustrative example in Section 2.1 to demonstrate the process of identifying and selecting platoon join and split nodes. Given the solo mode routes of vehicle 1 as 1→8→20 and vehicle 2 as 6→5→4→24→19 (**Figure 5(c)**), we use the segments of 8→20 from vehicle 1 and 4→24 from vehicle 2 for illustrative purpose. With nodes {8} and {4} as candidate join locations, **Algorithm 2** returns $N_{max} = 4$ additional join locations (nodes $\{2,3,9,10\} \in J$ in **Figure 5(d)**). Similarly, consider nodes {20} and {24} as candidate split locations and return 4 more split locations (nodes $\{16,21,22,23\} \in S$ in **Figure 5(d)**). By iterating the nodes in sets of $J$ and $S$, we can calculate the new cost of modified vehicle routes and find that platoon segment 9→21 generates the most savings from solo mode results (**Figure 5(e)**). However, vehicle 2 is still responsible to deliver requests 2 and 3 at this stage. According to **Algorithm 1**, we reconstruct the delivery paths of vehicle 1 and 2 after the platoon split and search for en-route transfers. In this case, we find that relocating request 2 from vehicle 2 to vehicle 1 over the platoon segment 9→21 could reach even more cost savings than the solo mode (**Figure 5(f)**).

## 4.2 Multi-vehicle platoon joining

In previous section, we propose a customized neighborhood search algorithm to find the minimal platoon formation consisting of only two vehicles. To search and construct multi-vehicle platoons, **Algorithm 3** first iteratively merges platoons together from the results of Section 4.1, and then insert remaining individual vehicles into feasible platoons found previously to explore all the join possibilities.

**Algorithm 3.** Multi-vehicle platoon search

| | |
|---|---|
| Input: | Algorithm 1 inputs and outputs, maximum platoon length $u$, $\varsigma = 1$. |
| 1. | **While** $\varsigma = 1$ **do** |
| 2. | **For** any $p_i^M, p_j^M \in P^M$ **do** |
| 3. | Identify the *LCPS* $l_i$ in $p_i^M$ and $l_j$ in $p_j^M$. |
| 4. | **For** each segments in $l_i$ and $l_j$ **do** |
| 5. | Identify the *LCPS* $l_{ij}$ between selected segments of $l_i$ and $l_j$. |
| 6. | **If** $l_{ij}$ exists **then** |
| 7. | Merge $p_i^M$ and $p_j^M$ over $l_{ij}$, and calculate new costs while satisfying $u$. |
| 8. | Search for platoon extension and en-route transfers (as *Algo 1, lines 8-14*). |
| 9. | Rank and select the platoon joins with most savings than $C^M$ and update $P^M$ as $P^{M'}$. |
| 10 | **If** no platoon join can be found **then** $\varsigma = 0$ |
| 11. | **For** any $\{k \in K | k \notin P^{M'}\}$ and $p_m^M \in P^{M'}$ **do** |
| 12. | Insert $p_k^S \in P^S$ into $p_m^M$ and calculate the new cost while satisfying $u$. |
| 13. | Search for platoon extension and en-route transfers (as *Algo 1, lines 8-14*). |
| 14. | Rank and select the individual vehicle insertions into $P^{M'}$ with costs $C^{M'}$ and update $P^S$ as $P^{S'}$ with costs $C^{S'}$. |
| Output: | Solo mode $P^{S'}$ with $C^{S'}$ and modular mode $P^{M'}$ with $C^{M'}$. |

Note: *LCPS* = longest common platoon segments

We keep searching for join between platoons while there still exists any platoon that has not been explored yet or new platoon join is just created. If any new platoon is created, we re-iterate the join between all platoons again. At each iteration, by choosing any pair of platoons, we first identify the longest common platoon segments (LCPS) for each platoon group (lines 2 to 3). Then, the heuristic iterates over each segment between the two LCPS and search for their common platoon paths again (lines 4 to 5). Note that our proposed heuristic algorithm solves the MDARP based on shortest-path distance and time, which might leave out intermediate nodes along the routes and leads to difficulty in identifying the LCPS. To tackle this, all the nodes along the shortest path of each segment are listed and then used to identify the LCPS. Once a common platoon path is found, the heuristic merges these two platoons together over the shared segments and re-calculates the new cost to ensure the capacity constraints and platoon length limit are satisfied (lines 6 to 7). Then, similar to **Algorithm 1**, we continue to search for platoon extension and passenger en-route transfers (line 8). If there is no further possibility to merge between platoons, remaining individual vehicles are iteratively inserted into existing platoons (lines 11 to 13). **Algorithm 3** returns the vehicle routes for modular mode and remaining unmatched individual vehicle routes from solo mode.

The overall heuristic with all the components is summarized in **Figure 6**.

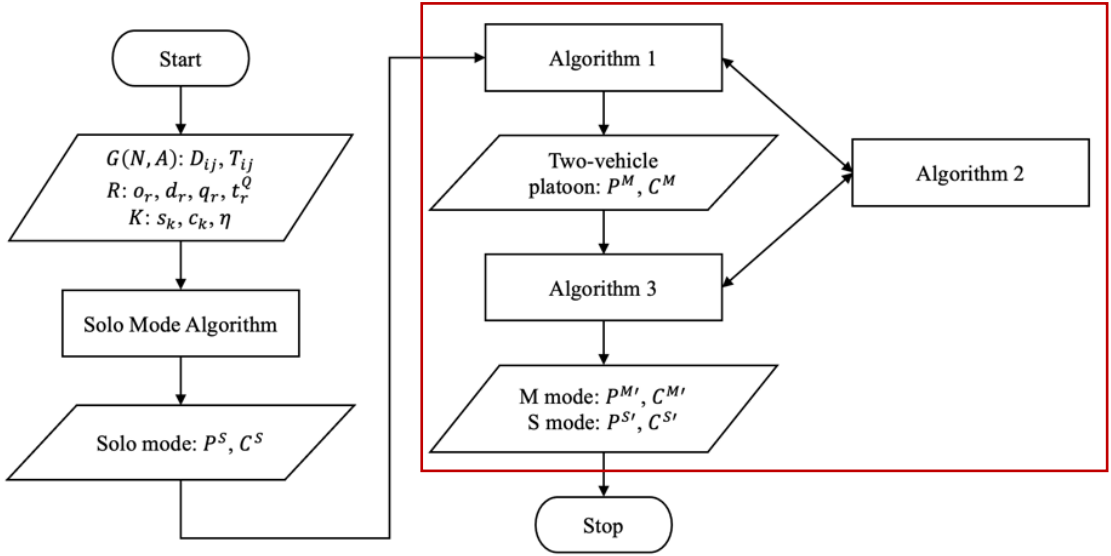

Figure 6. Flow diagram of overall heuristic algorithm (in the red box) integrated with a DARP algorithm.

## 5. Numerical experiments

Numerical experiments are conducted in this section to (1) evaluate the computational performance and (2) explore the potential benefits of modular vehicle technology. We first implement small-scale tests on a simple network. The optimal solutions obtained from the MILP model and the corresponding computation times are compared with our proposed heuristic algorithm. To solve the MILP model, we used Gurobi 8.1.1 optimization software as the commercial solver, running on a 64-bit Windows 8.1 personal computer with the Intel Core i7-6700K CPU and 40 gigabyte RAM. To further explore the potential benefits and optimal operation scenarios of MVs, large-scale and more practical instances are tested on the Anaheim network (378 nodes and 796 arcs, after removing centroids) with our proposed heuristic algorithm. The network information can be found in Github (Stabler, 2022). For convenience, the initial feasible solo mode solutions are obtained using a basic insertion heuristic (see Fu and Chow, 2022) without any loss of generality in the effectiveness of the proposed heuristic.

### 5.1 Small-scale test
The goal of the small-scale test is to evaluate the computational efficiency of a commercial solver for the MILP model and the heuristic, as well as the optimality of the heuristic algorithm under different scenarios.

Two operation policies, solo mode and modular mode, are compared. The optimal solutions found using a benchmark MILP commercial solver with branch-and-bound/-cut methods are compared with our heuristic algorithm. Since the MDARP with different vehicle and passenger locations requires enormous computation power to solve exactly, we are only able to handle examples involving at most 4 vehicles and 4 requests within 2 hour computation time limit.

Three scenarios ($S$) are tested on the network shown in **Figure 7**. Each scenario has a different combination of vehicles and requests: ($S1$) $|K| = 2, |R| = 3$, ($S2$) $|K| = 3, |R| = 4$, and ($S3$) $|K| = 4, |R| = 4$. There are 13 nodes and 40 links on the network, where the travel cost is labeled next to each link in both directions. For the MILP model, two layers are used. For convenience, we assume that the travel distance is in units of miles and the vehicle travel speed to be 1 mile/min for all tests. As a result, the travel cost in distance and time are in the same value for each link. Small-scale test instances, including vehicle initial locations, passenger pick-up and drop-off locations, and number of passengers in each request, can be found on Github (Fu, 2022). The vehicle capacity is set to be 4 and the request in-system time is set to be 0 in all small-scale tests. Various combinations of objective weight values are chosen to evaluate the trade-offs between operator cost and passenger cost. Since request on average consists of more than 2 passengers, the weight values vary from $\alpha:\beta = 1:1$, $\alpha:\beta = 2:1$, $\alpha:\beta = 4:1$, $\alpha:\beta = 6:1$ to $\alpha:\beta = 1:0$. In addition, we set the platoon saving rate $\eta$ to be 5% and 10%.

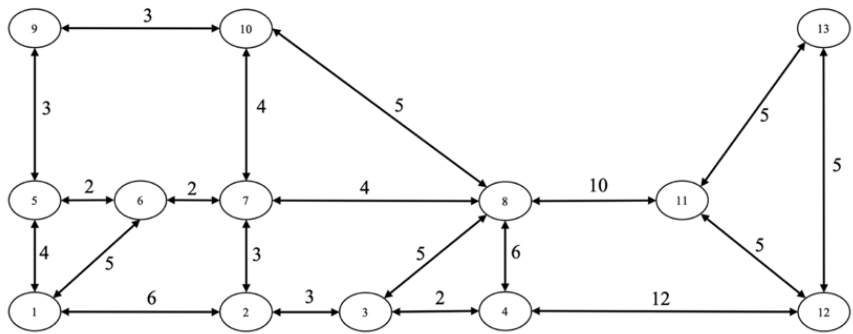

Figure 7. Small-scale test network.

The computation accuracy is summarized in **Figure 8** for various objective weight values. We consider the optimal objective value obtained from solo mode and solved from the MILP model as the baseline (the 100%, light grey column without shading in **Figure 8**). The optimality gap between MILP solution and heuristic algorithm (HA) can be observed by comparing them for the same scenario respectively: solo mode, modular mode with $\eta = 5\%$, and modular mode with $\eta = 10\%$.

HA performs best when considering passenger service time in the objective function (average of 0.17% when $\beta = 1$). Overall, our proposed HA only shows an average of 0.6% optimality gap from the objective value obtained by MILP solver. The cost savings and benefits from MVs can be observed by comparing the MILP objective values across different scenarios. Without considering the passenger service time, MV saves about 10% when $\eta = 5\%$ and about 15% when $\eta = 10\%$ against the solo mode. When the cost on passenger side is also included, the savings from MV mode decrease to about 4.1% when $\eta = 5\%$ and 6.6% when $\eta = 10\%$. From all observations, we find that MVs save more cost when the operator side gets more attention and weight, i.e. passengers are more inelastic to travel cost.

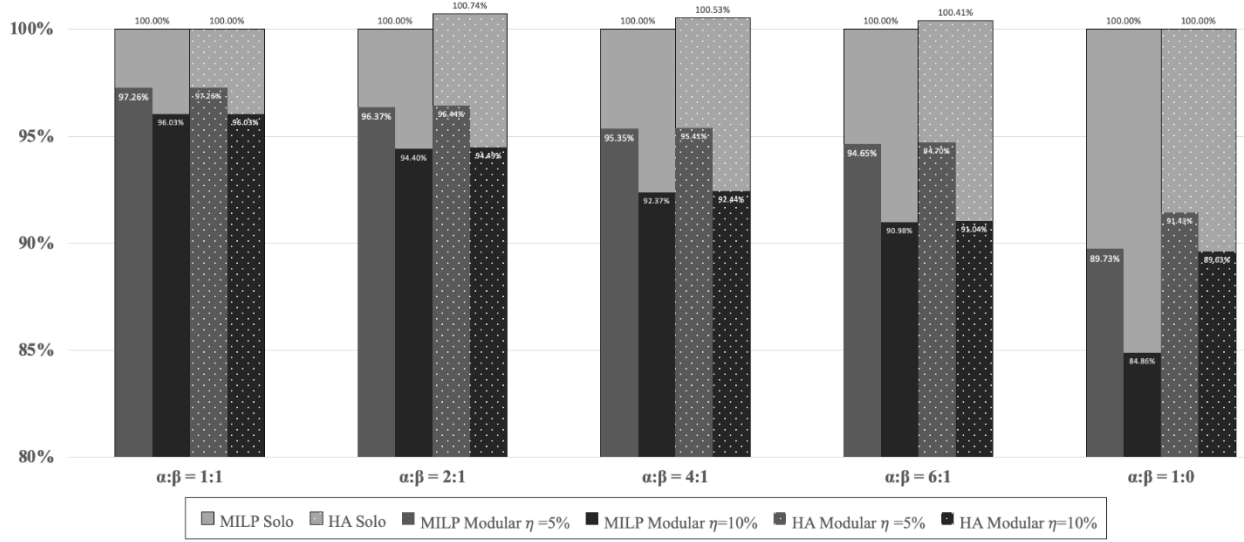

Figure 8. Comparison of objective values under different operations and solution algorithm.

The computation time for solving MILP model and heuristic algorithm under different operation modes and scenarios are summarized in **Table 4(a)** and **Table 4(b)**. The required computation time of the MILP model increases dramatically with the consideration of passenger service time under MV mode, whereas our proposed heuristic algorithm remains stable for all cases. From **Table 4(b)**, the computation time of MILP model increases by more than 10 times over the three scenarios with different vehicle and request numbers. By contrast, the heuristic algorithm only requires a few seconds before obtaining the solution and outperforms the MILP solver as the problem size increases. The problem size in this small-scale test only goes from $|K| = 2, |R| = 3$ in $S1$, $|K| = 3, |R| = 4$ in $S2$, to $|K| = 4, |R| = 4$ in $S3$ on a 13-node, 40-link network. The results from **Table 4** show that the MILP model is not practical for realistic scenarios. Overall, our proposed heuristic algorithm can reach an average optimality gap of 0.57% with only a fraction of the required computation time against the MILP model from the set of small experiments.

Table 4. Computation time over various scenarios
(a) Objective weights $\alpha$ and $\beta$

| Method | Mode | $\alpha:\beta = 1:1$ | $\alpha:\beta = 2:1$ | $\alpha:\beta = 4:1$ | $\alpha:\beta = 6:1$ | $\alpha:\beta = 1:0$ |
|---|---|---|---|---|---|---|
| MILP | Solo | 7.03 | 4.31 | 3.08 | 1.75 | 0.08 |
|  | Modular $\eta = 5\%$ | 2748.98 | 1752.21 | 720.16 | 696.47 | 20.01 |
|  | Modular $\eta = 10\%$ | 2671.54 | 2097.99 | 1120.92 | 755.44 | 12.53 |
| HA | Solo | 0.01 | 0.01 | 0.01 | 0.01 | 0.01 |
|  | Modular $\eta = 5\%$ | 15.53 | 15.36 | 14.36 | 14.61 | 13.58 |
|  | Modular $\eta = 10\%$ | 15.40 | 15.07 | 14.56 | 14.45 | 13.91 |

(b) Vehicle and request numbers

| Method | Mode | $S1$ | $S2$ | $S3$ |
|---|---|---|---|---|
| MILP | Solo | 0.86 | 5.11 | 3.78 |
|  | Modular $\eta = 5\%$ | 5.41 | 332.78 | 3224.51 |
|  | Modular $\eta = 10\%$ | 8.09 | 226.47 | 3760.50 |
| HA | Solo | 0.00 | 0.00 | 0.00 |
|  | Modular $\eta = 5\%$ | 9.58 | 12.50 | 21.98 |
|  | Modular $\eta = 10\%$ | 9.62 | 12.39 | 22.02 |

Note: all computation times are shown in seconds

## 5.2 Large-scale tests

To explore the potential benefits and identify the ideal operation scenarios of modular vehicle technology, we apply our heuristic algorithm to instances on the Anaheim network, as shown in **Figure 9** with 378 nodes and 796 arcs (zone centroids 1-38 and linked arcs are removed from the network), to conduct more realistic large-scale numerical experiments. All 378 nodes are considered potential join/split nodes as well pickup or drop-off locations. The shortest-path travel distance and travel time matrices are calculated beforehand and imported at the beginning of our algorithms for each pair of nodes. The unit of travel distance on each link is in miles and the unit of free-flow travel time on each link is in minutes.

Six random instances are drawn each from a set of 40 scenarios (10 different problem sizes and 4 different spatial distributions) resulting in a total of 240 instances, from which the output measures are used to fit a linear regression model to characterize the relationship of the different parameters on the cost reduction from solo mode.

The large-scale experiment settings are summarized in **Table 5**. A total number of ten different problem sizes are defined ranging in terms of the number of vehicles and requests from $K = 5, R = 8$ to $K = 25, R = 50$. For each size, we consider four spatial distribution scenarios for the vehicle initial locations, request pickup and drop-off locations: 1) uniformly randomly generated over the network; 2-4) clustered around randomly selected node with 3, 5, or 10 neighborhoods, respectively.

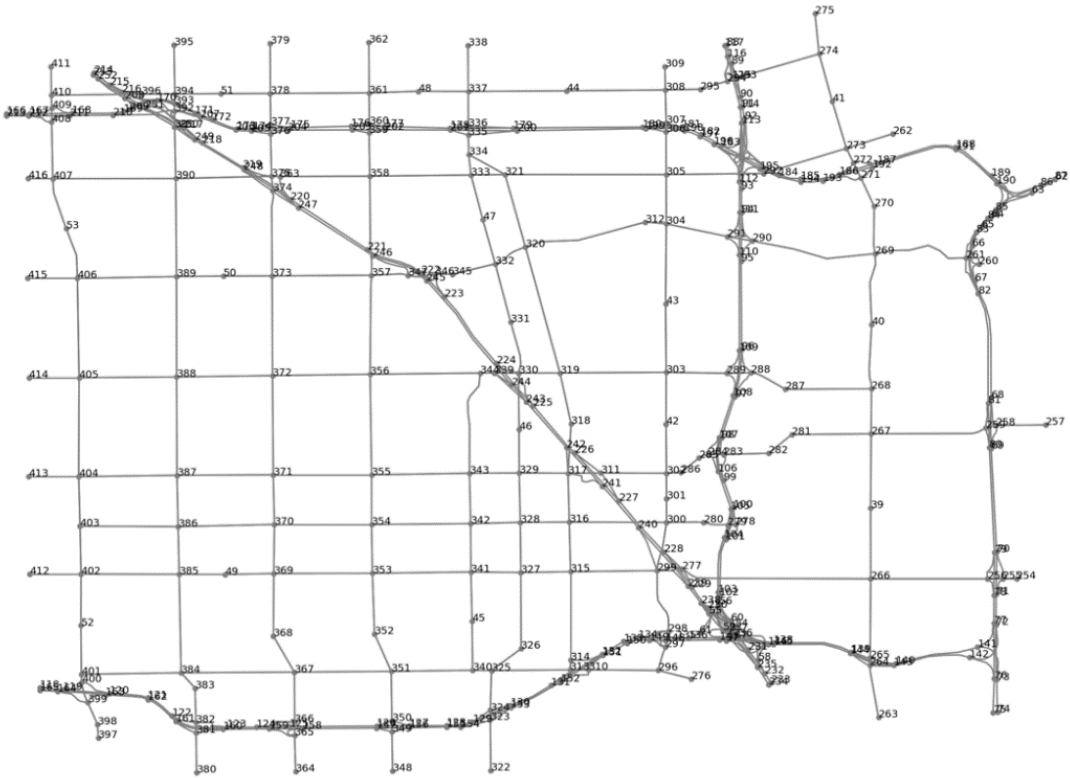

Figure 9. Anaheim network with 378 nodes and 796 arcs.

Table 5. Large-scale experiment settings

| *Instance-specific* | |
|---|---|
| Vehicle and request numbers $(K, R)$ | (5,8), (5,10), (10,15), (10,20), (15,20), (15,30), (20,30), (20,40), (25,40), (25,50) |
| Objective weights | $\alpha:\beta = 1:1$ or $\alpha:\beta = 3:1$ |
| Maximum platoon length | Randomly selected in [4,5,6,7] |
| Vehicle capacity | Randomly selected in [4,5,6,7] |

| Platoon saving rate ($\eta$) | Randomly selected in [5%, 6%, 7%, 8%, 9%, 10%] |
|---|---|
| *Individual-specific* | |
| Vehicle and request locations | 1) Uniformly randomly generated over the network (U) |
| (Spatial distribution) | 2) Clustered around randomly selected 3 neighbors (C3) |
| | 3) Clustered around randomly selected 5 neighbors (C5) |
| | 4) Clustered around randomly selected 10 neighbors (C10) |
| Request in-system times | 1) All at $T = 0$ |
| (Temporal distribution) | 2) Uniformly generated over [0,1] |
| | 3) Uniformly generated over [0,4] |
| Passenger number in each request | Randomly selected in [1,2,3,4] |

### 5.2.1. Summary of results of the 240 samples

The other parameters are randomly generated among the each of the 40 scenarios to build the 240 samples. For the temporal distribution, we assume all vehicles are ready for service at $T = 0$. However, each request might enter the system and ask for service at different times: 1) all requests are ready at $T = 0$; 2-3) uniformly randomly generated over [0,1] or [0,4] mins. For the objective weight values between vehicle travel distance and request service time, we randomly select between $\alpha:\beta = 1:1$ and $\alpha:\beta = 3:1$. The maximum allowed platoon length and vehicle capacity limit are randomly selected among [4,5,6,7]. For the MV platoon saving rate $\eta$, we randomly select among [5%, 6%, 7%, 8%, 9%, 10%]. The number of passengers in each request is randomly selected from [1,2,3,4]. The set of 240 generated instances can be found on Github (Fu, 2022).

Computation times for solo (S) mode and modular (M) mode over the ten problem sizes (in terms of vehicle and request numbers) are shown in **Figure 10**. The required computation time for solo mode remains within 10 seconds for all tests. As for modular mode, the computation time steadily increases with the problem size while our proposed heuristic algorithm can still handle 25 vehicles and 50 requests ($K = 25, R = 50$) within 1 hour. By contrast, the MILP model supported by commercial solver Gurobi can barely deal with 4 vehicles and 4 requests on a 13-node network within 2 hours.

The impact of spatial and temporal distribution of vehicles and requests on various performance metrics have been summarized in **Table 6**. The percentages presented in the rows of vehicle travel cost, request service time and total cost of **Table 6** are computed as follows,

$$percentage\ of\ cost\ differences = \frac{Modular\ mode - Solo\ mode}{Solo\ mode} \times 100\%,$$

where a negative sign means that the modular mode performs better and the cost is less than the corresponding solo mode. From the large-scale results, we find that more clustered spatial distribution would lead to more cost savings in all aspects: vehicle travel distance, request service time, and total cost. It is also more likely to have increased platoon numbers, more en-route transfers, higher involvement of platooning vehicles, and longer platoon sizes. As for the temporal distribution, denser demand would lead to higher savings in vehicle travel cost and longer platoon sizes. However, it is preferred to have low demand density from the perspective of request service time, total cost and en-route transfer opportunities.

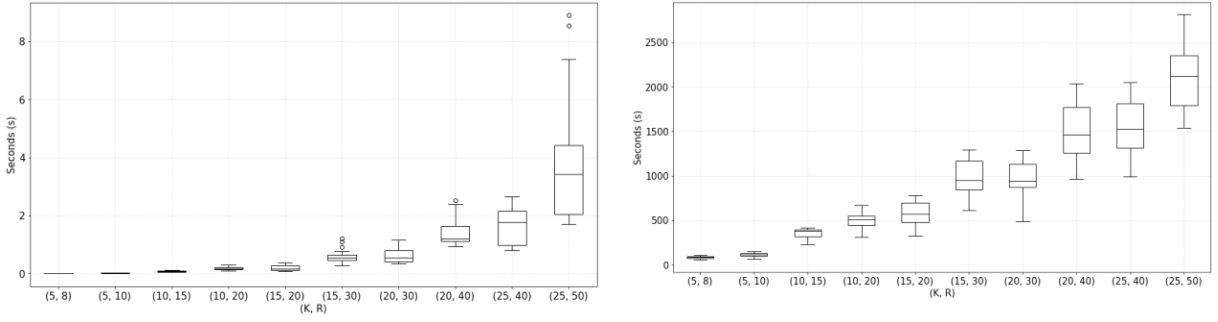

(a) Solo mode            (b) Modular mode

Figure 10. Computation time of solo and modular mode.

Among all randomly generated large-scale instances on the Anaheim network, the highest savings can reach up to 52.04% in vehicle travel cost, 35.58% in request service time, and 29.40% in total cost (**Table 6(a)**), which reveals the promising potentials of modular vehicle technology even under practical scenarios.

Table 6. Average performance metrics

(a) Summary statistics

| Percentage of cost differences | Mean | Standard deviation | Min | Max |
|---|---|---|---|---|
| Vehicle travel cost | -6.94% | 11.43% | -52.04% | 18.78% |
| Request service time | -1.81% | 3.9% | -35.58% | 3.51% |
| Total cost | -3.02% | 4.13% | -29.4% | 0% |

Note: Number of observations = 240

(b) Average performance metrics grouped by spatial or temporal distribution

| Metrics | Spatial distribution | | | | Temporal distribution | | |
|---|---|---|---|---|---|---|---|
| | U | C10 | C5 | C3 | [0,4] | [0,1] | $T=0$ |
| Vehicle travel cost | +0.01% | -4.00% | -7.04% | -16.48% | -6.75% | -6.53% | -7.57% |
| Request service time | -0.49% | -1.74% | -2.47% | -2.49% | -2.78% | -1.52% | -1.06% |
| Total cost | -0.43% | -2.35% | -3.56% | -5.64% | -3.80% | -2.75% | -2.44% |
| Number of platoons (per 100 vehicles) | 4.33 | 15.78 | 15.78 | 15.74 | 12.69 | 13.7 | 12.38 |
| En-route transfers (per 100 requests) | 1.58 | 2.09 | 1.96 | 2.24 | 2.76 | 1.38 | 1.77 |
| Vehicles in platoon (%) | 8.67% | 43.56% | 49.22% | 60.55% | 40.24% | 41.28% | 40.25% |
| Platoon size (avg.) | 2 | 2.76 | 3.12 | 3.85 | 3.17 | 3.01 | 3.25 |

### 5.2.2. Quantifying effects of spatial/temporal clustering with regression

To further explore the ideal operation scenarios and identify the key parameters that impact the efficiency of modular vehicles, we estimated a multiple linear regression on the 240 samples to quantify the effects of different parameters on the total cost differences between solo and modular mode (dependent variable), where a more negative difference means modular improves more in cost. The overall regression results are summarized in **Table 7(a)**. Detailed individual contributions of statistically significant predictors are summarized in **Table 7(b)**, where $Spatial\_C10$, $Spatial\_C5$, and $Spatial\_C3$ are binary independent variables and when $Spatial\_C10 = 0$, $Spatial\_C5 = 0$, and $Spatial\_C3 = 0$, it refers to the uniform spatial distribution. From the multiple linear regression results, we find that more clustered spatial distribution lead to more savings in total cost, whereas larger vehicle capacity would account for less total cost savings. On the contrary, we do not find statistically significant effects from temporal clustering, maximum platoon length, instance size, or saving rates on the cost savings relative to solo mode. This suggests that scaling to larger instances would not necessarily benefit or worsen platooning operations.

Table 7. Multiple linear regression results
(a) Model regression statistics

| Multiple R | $R^2$ | Adjusted $R^2$ | Standard Error | Observations |
|---|---|---|---|---|
| 0.5 | 0.25 | 0.23 | 0.04 | 240 |

(b) Detailed independent variables

| Independent Variables | Coefficients | Standard Error | t Stat | P-value |
|---|---|---|---|---|
| Spatial_C10 | $-1.73 \times 10^{-2}$ | $6.66 \times 10^{-3}$ | -2.59 | $1.01 \times 10^{-2}$** |
| Spatial_C5 | $-3.03 \times 10^{-2}$ | $6.61 \times 10^{-3}$ | -4.59 | $7.29 \times 10^{-6}$*** |
| Spatial_C3 | $-5.05 \times 10^{-2}$ | $6.59 \times 10^{-3}$ | -7.66 | $4.86 \times 10^{-13}$*** |
| Veh_Cap | $7.21 \times 10^{-3}$ | $2.17 \times 10^{-3}$ | 3.33 | $1.01 \times 10^{-3}$*** |
| Constant | $-4.59 \times 10^{-2}$ | $1.33 \times 10^{-2}$ | -3.44 | $6.85 \times 10^{-4}$*** |

Note: *** Significant at 0.01; ** Significant at 0.05; * Significant at 0.1

The results in **Table 7** confirm that clustered trip patterns are more conducive to modular platooning benefits, which suggest that platooning can be applicable to providing service to dense residential areas to connect them to the local CBD. In such operations, vehicles can operate in solo mode to collect and distribute passengers and platoon together to take the long-haul trip. This type of operation is labeled as a hub-and-spoke design for MVs in Caros and Chow (2020), where it is shown to be the best design for maximizing consumer surplus compared to other operations like door-to-door microtransit or running only first/last mile access. In other words, MVs can maximize their platooning benefits by identifying major residential neighborhoods or enclaves and serving those via hub-and-spoke to the CBD and vice versa.

### 5.2.3. *Illustration of platooning under different spatial clustering*
Two large-scale examples, both with 25 vehicles and 50 requests, are visualized in **Figure 11(a)** with spatial distribution $C3$, and **Figure 11(b)** with spatial distribution $C10$. Vehicle initial locations, request origins and destinations, platoon join and split locations are marked on the Anaheim network. For simplicity, we only highlight the paths where modular vehicles travel in platoon.

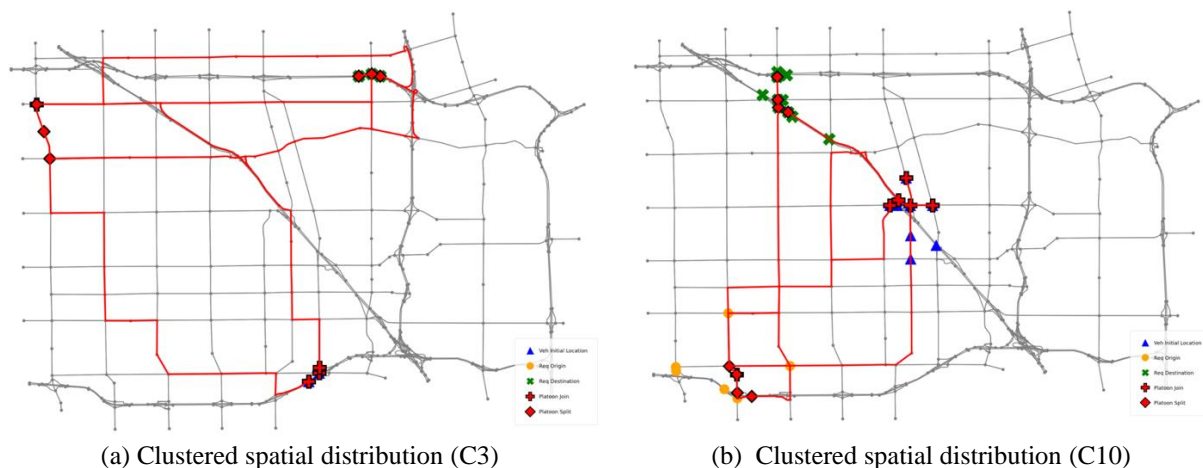

(a) Clustered spatial distribution (C3)     (b) Clustered spatial distribution (C10)
Figure 11. Large-scale experiment examples: $K = 25\ R = 50$.

## 6. Conclusion
With the ability to connect and disconnect as platoons and thus expand and reduce its capacity limit according to demand, MVs have the potential to reduce both operator cost and passenger costs in demand-responsive transit, the so-called dial-a-ride problem. The concept of MVs has attracted much research attention in recent years. However, all relevant studies on the innovative modular vehicle technology focus on either vehicle platooning problem, pickup and delivery problem with transfers, or modular vehicle with variable capacity on public transit services, separately. None of existing literature has addressed all these three major challenges together for the operation of MVs with modular platooning and en-route transfers.

In order to find the optimal assignments and routes of MVs, we first formulate a MILP model for the MDARP that integrates vehicle platooning with request pickup and delivery, considers passenger en-route transfers during vehicle platooning, and addresses the variable capacity feature of platoons at the same time. Two weighted objective values of vehicle travel cost and passenger service time are optimized in the MILP model. The vehicle docking and undocking process for platoons as well as the passenger transfers are strictly synchronized between vehicles on both temporal and spatial dimensions. The model requires an undirected graph structure with multiple layers.

Since the MDARP can be simplified to a DARP which is known to be NP-hard, a Steiner tree-inspired local neighborhood search algorithm is proposed to solve large-scale problems. Based on initial feasible solo mode solutions, the proposed heuristic algorithm consists of two major steps: 1) modify the solo mode solutions and find two-vehicle MV platoons, 2) merge between feasible MV platoons and then insert individual vehicles to platoons to find multi-vehicle platoons. To validate the performance of our proposed heuristic algorithm, small-scale experiments show that our proposed heuristic algorithm can reach an average optimality gap of 0.57% with only a fraction of the required computation time against the MILP model. To further explore the potential benefits and identify the ideal operation scenarios of MVs, a set of large-scale experiments are implemented on a 378-node Anaheim network. To understand the role of different factors in benefits for platooning, we randomly construct 240 large-scale instances on this network and estimate a linear regression model with the output data. Results reveal that more clustered spatial distribution and smaller vehicle capacity would lead to more savings in the total cost, while such factors as temporal clustering, maximum platoon length, instance size, and saving rates are not statistically significant. Depending on the operational settings, using MVs can save up to 52% in vehicle travel cost, 36% in passenger service time, and 29% in total cost against existing mobility-on-demand services.

There are several future research directions that we can continue our work on. First of all, the operation performance of MVs needs to be tested and verified under more dynamic, uncertain and realistic settings with simulation-based evaluation methods. Second, since the modular vehicle concept is most likely deployed with electric vehicles, operators might need to consider the charging process in the operation of MVs. Third, since MVs are physically connected with each other, more innovative applications could be further explored based on features of MVs, such as mobile charging-as-a-service (Abdolmaleki et al., 2019), and integrated service of logistics and passengers (Hatzenbühler et al., 2022).

## ACKNOWLEDGMENTS
The authors are partially supported by the C2SMART University Transportation Center (USDOT #69A3551747124) and NSF CMMI-2022967. The extended abstract of this paper presented at the 15th International Conference on Advanced Systems in Public Transport (CASPT 2022).

## AUTHOR CONTRIBUTIONS
All authors, ZF and JYJC, confirm contributions to the study conception and design, analysis and interpretation of results, and manuscript preparation of the paper. All authors reviewed the results and approved the final version of the manuscript.